# Doubly-charged negative ion of $C_{60}$ molecule

A. Z. Msezane[1] and A. S. Baltenkov[2]

[1]Center for Theoretical Studies of Physical Systems,
Clark Atlanta University, Atlanta, Georgia 30314, USA
[2]Institute of Ion-Plasma and Laser Technologies
Tashkent 100125, Uzbekistan

**Abstract.** Within the Dirac- and Lorentz-bubble potential models an electronic structure of the doubly-charged negative ion $C_{60}^{2-}$ has been studied by a variational method. It is shown that even in the first approximation of this method when a trial wave function of the two electrons is represented as a product of one-electron functions the total energy of the system is negative, a manifestation of the existence of a stable state of the doubly-charged negative ion in these models. The second electron affinity of $C_{60}$ according to estimation is about $\varepsilon_2 \approx 1$ eV. The photodetachment cross sections $\sigma(\omega)$ of this ion have been calculated as well. Near threshold $\sigma(\omega)$ is found to exhibit peculiar and interesting behavior. The first cross section accompanied by the transformation of the doubly-charged negative ion into a singly-charged one is exponentially small near the process threshold. The second cross section corresponds to the photodetachment of a singly-charged ion; it increases at the threshold as a power function of the kinetic energy of the photoelectron. These cross sections are of the same order as the photodetachment cross sections of atomic ions with the same electron affinity.
**AMS (MOS) Subject Classification.** 70G75, 81V10, 81V55

## 1. Introduction

The photoionization of an atom A inside the endohedral anions: $A@C_{60}^{z-}$ was discussed in [1, 2]. In those calculations the charge of the negative molecular ion $z$ was varied within the range $0 \leq z \leq 5$. However, the question of the encapsulation of the atom A inside the hollow interior of the multiply-charged negative ions of $C_{60}$ was not addressed and the existence of the $A@C_{60}^{z-}$ systems was merely postulated in [1, 2]. Atomic multiply-charged ions (even for $z$=2) are, as a rule, unstable systems with the lifetimes $10^{-6}$–$10^{-7}$ s [3]. Their instability is due to the Coulomb repulsion of the extra electrons whose wave functions have the parent atomic nucleus as the center. Because of the strong overlap of electron densities the Coulomb interaction energy dominates the binding energy of the electrons with the atom, leading to the system's decay. Conceivably, under appropriate conditions the existence of multiply-charged anions $C_{60}^{z-}$ could be more favorable. Furthermore, it could also be imagined that the extra electrons (for example, for $z$=2) are localized mainly at the opposite sides of the fullerene sphere to minimize the Coulomb repulsion. Since the sphere radius is large on the atomic scale the repulsion energy could be insufficient to cause decay. The energy, $E$ of $z$-electrons confined within the short-range potential well $U(r)$ of the $C_{60}$ shell and forming the multiply-charged ion $C_{60}^{z-}$ is negative, i.e. $E<0$. On the other hand the Coulomb repulsion energy, $Q$ of these electrons is positive, namely $Q>0$. A stable bound state of the electrons in the potential well is possible if their total energy is $E + Q < 0$. In the first approximation the energies $E$ and $Q$ can be estimated as follows. The binding energy of one electron with $C_{60}$ is $\approx -2.7$ eV [4]. The upper limit of the binding energy of $z$ electrons in the well is $E \approx -2.7 \times z$ eV [*]. The Coulomb repulsion energy of the pair of electrons located in the potential well of radius $R = 3.55$ Å [4] has the value of about $e^2/R \approx 4.1$ eV. The number of such pairs is a binion $p = z!/[2(z-2)!]$. We now have the estimate for the total energy of the electrons $E + Q \approx -2.7z + 4.1p$. Substituting $z = 5, 4$ and $3$ in this formula, we conclude that the total energy of penta-, tetra- and triply-charged negative ions of $C_{60}$ is positive. Consequently, it is unlikely they can exist in a stable state. However, the total energy is negative for $z=2$ only.

In this context it is interesting and informative to investigate the possibility of the existence of stable doubly-charged negative ions of $C_{60}$ and, if they do exist, to determine their electronic structure. The system under consideration is similar to a helium-like positive ion with the only difference being that the coupling of each electron of the system is due to the short-range potential $U(r)$ rather than the long-range interaction of electrons with the positive nucleus. Therefore, to analyze this problem we apply the methods used for helium-like positive ions [5,6], namely: at first we consider the Coulomb interaction between electrons as a perturbation and estimate the binding energy of extra electrons in doubly charged ions $C_{60}^{2-}$, and then with the help of variational principle we will calculate the binding energy of the anion ground state.

In Ref.[7] for the doubly-charged carbon cluster anions $C_{60}^{2-}$ a semi-empirical approach to the problem of estimating the second electron affinity of $C_{60}$ was proposed. This approach is based on the assumption that the interaction of extra electrons with the fullerene shell could be described by some spherically symmetric potential $U(r)$ whose parameters are such that upon solving the wave equation for one electron in this potential well, the first electron affinity of $C_{60}$ molecule $\varepsilon_1$ is in agreement with the measured value of $\varepsilon_1 = 2.65 \pm 0.02$ eV. The solutions of the wave equation for a pair of extra electrons in the potential well $U(r)$ define the total energy of system and the value of the second electron affinity $\varepsilon_2$.

In this paper we use this approach for estimating the second electron affinity of $C_{60}$ applying the variational method. We will consider two types of the potentials simulating a potential of the fullerene shell $U(r)$ and calculate the wave functions of the single electron in these potentials (Sec. 2). Further, we will choose a shape of the trial wave function and calculate the total energy of the pair of electrons locked within these potential wells (Sec. 3). Varying the wave function parameters (rather than the fullerene radius $C_{60}$ as in paper [7]), we will find a minimum of the total energy of electrons and calculate the eigen values and eigen functions of the system. That will give an estimate of the second electron affinity of $C_{60}$ (Sec. 4). In Section 5 the calculated wave functions are used to calculate the photodetachment cross sections of the processes $C_{60}^{2-} + \hbar\omega = C_{60}^{-} + e$ and $C_{60}^{-} + \hbar\omega = C_{60} + e$ ($\hbar\omega$ is the photon energy). Section 6 gives the conclusions.

## 2. Potential of $C_{60}$ shell

For a shape of the $C_{60}$ shell potential to be selected, we will guide by the following requirements. The potential $U(r)$ is to be the attractive potential and in the potential well $U(r)$ an $s$-level with the binding energy equal to $E_s = -\varepsilon_1$ should exist. We will assume as in [7] that in the $1s^2$ ground state of the system $C_{60}^{2-}$ the both electrons are in state with zero orbital moment and antiparallel spins.

---

[*] According to [4], the extra electron is localized in the ground state of the $p$-like level. So the electron configuration $2p^z$ is quite acceptable. If one considers the ion ground state to be an $s$-like state, then the state $1s^2 2p^{z-2}$ will correspond to the minimal energy. Evidently, the absolute value of the energy of this configuration is less than the upper estimate.

The function $U(r)$ should be localized in a rather thin spherical shell with the width $\Delta$ of about few atomic units in the vicinity of the fullerene radius $R$ [8]. The potential $U(r)$ is to have the minimum at the radius $r=R$, i.e. the bottom of the potential well cannot be flat (for details see [9-11]). The calculations of the potential well shape with the use of the density of collectivized $2s2p$ electron of $C_{60}$ shell (given in Ref.[8]) show that the potential meeting all those requirements is very close to the Lorentz-bubble potential

$$U(r) = -\frac{U_L}{(r-R)^2 + d^2}, \qquad (1)$$

that we will call it by analogy with the Dirac-bubble potential [4,12]

$$U(r) = -U_D \delta(r-R). \qquad (2)$$

In Eq. (1) the thickness of the potential well $\Delta$ at the middle of the maximal depth is $\Delta=2d$ and $U_0=U_L/d$ is the maximal depth of the potential well (1). The parameters of potential (1) should be connected with each other in such a way that in the potential well (1) there is a $s$-like state with the specified energy $E_s = -\varepsilon_1$. These parameters are defined by the numerical solving of the wave equation for single electron moving in the potential well (1).

Potential (2) formally can be considered as the Lorentz-bubble potential (1) with zero-thickness $\Delta=0$ and infinity depth. The substitution of pseudopotential (2) in the wave equation provides a specified jump of the logarithmic derivatives $\Delta L$ of the wave function at the point $r=R$. In turn, the value $\Delta L = -2U_D$ is defined by the first electron affinity $\varepsilon_1$ of $C_{60}$. The solutions of the wave equation with the Dirac-bubble potential can be written analytically (for details see [4,12-14]). In future we will use the model potentials (1) and (2) for description of the electronic structure of $C_{60}^{2-}$ anion.

### 3. Mean value of energy of Coulomb interaction

Let us study the behavior of a pair of the electrons in the Dirac-bubble potential well (2). Neglecting the spin-orbital interaction, we will write the Hamilton operator of the system in the following form

$$\hat{H} = \hat{H}_0 + V, \qquad (3)$$

where

$$\hat{H}_0 = -\frac{1}{2}(\nabla_1^2 + \nabla_2^2) - U_D \delta(r_1 - R) - U_D \delta(r_2 - R) \qquad (4)$$

is the Hamilton operator of two free electrons in the potential well; $V = |\mathbf{r}_1 - \mathbf{r}_2|^{-1}$ is the operator of the Coulomb interaction between the electrons. Throughout the paper the atomic units (au) ($\hbar = m = |e| = 1$) are used. In the zero approximation (when the Coulomb interaction $V$ is neglected) the problem for the both electrons reduces to that in Refs. [4,12] considering the behavior of single electron in the Dirac-bubble potential. In this approximation each of the electrons with energy $E_s = -\beta^2/2 = -\varepsilon_1$ obeys the wave equation

$$\left[-\frac{1}{2}\nabla_{1,2}^2 - U_D \delta(r_{1,2} - R)\right]\psi(\mathbf{r}_{1,2}) = -\frac{\beta}{2}\psi(\mathbf{r}_{1,2}). \qquad (5)$$

Here the strength of the delta-potential is defined by the formula [13,14]

$$U_D = -\Delta L/2 = \beta(1 + \coth \beta R)/2. \qquad (6)$$

Thus, in the Dirac-bubble potential model (2) two experimentally observed parameters, namely the fullerene radius $R$ and the energy of the ground $s$-state (the first electron affinity), fully define equation (5) and hence the whole spectrum of electronic states of the negative ion $C_{60}$. The solutions of the wave equations for one electron with zero orbital moment $\psi_{1s}(\mathbf{r}_{1,2}) = [\chi_{\beta 0}(r_{1,2})/r_{1,2}]Y_{00}(\mathbf{r}_{1,2})$ can be represented as follows

$$\chi_{\beta 0}(r_{1,2}) = B\frac{\exp(-\beta R)}{\beta R}\sinh \beta r_{1,2} \qquad \text{for } r_{1,2} \leq R,$$

$$\chi_{\beta 0}(r_{1,2}) = B\frac{\sinh \beta R}{\beta R}\exp(-\beta r_{1,2}) \qquad \text{for } r_{1,2} \geq R, \tag{7}$$

where $B$ is the normalized factor

$$B = \frac{z\sqrt{\beta}\exp(z/2)}{(\sinh z + \cosh z - z - 1)^{1/2}}; \quad z = 2\beta R. \tag{8}$$

In the first approximation of the perturbation theory the energy of the ground state of the system is equal to $E_{tot} = 2E_s + Q$, where $Q$ is the mean value of the energy of the Coulomb repulsion of two electrons in the potential well

$$Q(\beta) = \iint \Phi_\beta^*(\mathbf{r}_1,\mathbf{r}_2)\frac{1}{|\mathbf{r}_1 - \mathbf{r}_2|}\Phi_\beta(\mathbf{r}_1,\mathbf{r}_2)d\mathbf{r}_1 d\mathbf{r}_2. \tag{9}$$

Here $\Phi_\beta(\mathbf{r}_1,\mathbf{r}_2) = \psi_{1s}(\mathbf{r}_1)\psi_{1s}(\mathbf{r}_2)$ is the wave function of two non-interacting electrons; the integration in (9) is made over the six-dimensional configuration space. When integrating (9) the Coulomb potential is represented as a series in spherical function $Y_{lm}(\mathbf{r}_{1,2})$. The wave functions $\Phi_\beta(\mathbf{r}_1,\mathbf{r}_2)$ are independent of angular variables. Therefore, for integration over spherical angles of vectors $\mathbf{r}_{1,2}$, in this series all the terms except those with $l = m = 0$ will be zero. Thus, the integral (9) is transformed into the form

$$Q(\beta) = \int_0^\infty \chi_{\beta 0}^2(r_1)W(r_1)dr_1 \tag{10}$$

where

$$W(r_1) = \left[\frac{1}{r_1}\int_0^{r_1}\chi_{\beta 0}^2(r)dr + \int_{r_1}^\infty \frac{\chi_{\beta 0}^2(r)}{r}dr\right] \tag{11}$$

is the mean potential created by the electron cloud $\psi_{1s}(\mathbf{r})$ at the point $\mathbf{r}_1$. For $R$=3.527 Å= 6.665 au [15] the numerical value of the energy of Coulomb interaction $Q(\beta)$ in the well with the binding energy of one electron $E_s$=-2.65 eV is: $Q(\beta)$=3.72 eV (In Sec. 1 this energy was estimated as 4.1 eV) while the energy of non-interacting electrons is: $2E_s$. Hence in the Dirac-bubble potential model for $C_{60}$ shell the total energy of the system $E$=-5.30+3.72=-1.58 eV is negative.

In the Lorentz-bubble potential model the one-electron wave functions are the solutions of the wave equation with potential (1). These wave functions for the different thicknesses of potential wells were calculated by the Runge-Kutta method [16]. The results are presented in Fig. 1. For comparison, the wave function calculated with the Dirac-bubble potential ($\Delta = 0$) is given in the same figure. With the rise in the parameter $\Delta$ the cusp-behavior of the wave function for zero-thickness changes to more smoothly behavior near the point $r \approx R$. According to Fig. 1, the shape of the wave functions depends comparatively weakly on the parameters of the potential wells in which the electron is localized.

Let us use these wave functions to calculate with the formula (9) the energies of the Coulomb interaction $Q(\beta)$ of electrons and their total energy. The calculation results for different parameters of the potential wells are given in Table 1.

Table 1. Total energy of electrons in the potential wells (1) and (2)

| $\Delta$, au | $U_0$, au | $Q(\beta)$, eV | $E_{tot}$, eV |
|---|---|---|---|
| 0 | $\infty$ | 3.72 | -1.58 |
| 1 | 0.4415 | 3.68 | -1.62 |
| 2 | 0.2805 | 3.65 | -1.65 |
| 3 | 0.2243 | 3.62 | -1.68 |

A decrease in the energy of the Coulomb interaction $Q$ with the rise in $\Delta$ is quite explainable. The regions of electron delocalization increase and therefore the overlapping integrals of the wave functions (10) and (11) decrease. The total energy of electrons $E_{tot}$, locked in all the potential wells (1) is negative, so the existence of the double-charged negative ion of $C_{60}$ within the Lorentz- or Dirac-bubble potential models is quite acceptable. The second electron affinity $\varepsilon_2$ is the energy necessary to remove one of the extra electrons from the system and it is equal to the difference $\varepsilon_2 = E_{tot} - E_s$; the numerical value of this energy according to Table 1 is about 1 eV.

### 3. Variational method in the Dirac-bubble potential well

More exact values of energy and a wave function of the ground state of the double-charged negative ion $C_{60}$ can be obtained by a direct variational method. We replace the wave vector $\beta$ by $\gamma$ in the wave function $\Phi_\beta(\mathbf{r}_1, \mathbf{r}_2) = \psi_{1s}(\mathbf{r}_1)\psi_{1s}(\mathbf{r}_2)$ and in the one-electron functions $\psi_{1s}(\mathbf{r}_{1,2})$ and consider the wave vector $\gamma$ as a variational parameter. The problem of finding the energy of the ground state of two electrons in the Dirac-bubble potential well reduces to calculating the following integral

$$E(\gamma) = \iint \Phi_\gamma^*(\mathbf{r}_1, \mathbf{r}_2) \hat{H} \Phi_\gamma(\mathbf{r}_1, \mathbf{r}_2) d\mathbf{r}_1 d\mathbf{r}_2, \qquad (12)$$

and finding a minimum of the function $E(\gamma)$. The Hamilton operator of the system in (12) is defined by formulas (3) and (4). As a result we obtain the following expression for the total energy of the system as a function of the wave vector $\gamma$

$$E_{tot}(\gamma) = -\gamma^2 + 2[U_D(\gamma) - U_D]|\chi_{\gamma 0}(R)|^2 + Q(\gamma) \qquad (13)$$

where $Q(\gamma)$ is defined by formulas (10) and (11); $U_D(\gamma)$ is the function

$$U_D(\gamma) = \frac{\gamma}{2}(1 + \coth \gamma R). \qquad (14)$$

Another parameter $U_D$ is defined by the formula (6) and its numerical value is: $U_D \approx 0.44265$. Until now we impose no limitations on the vector $\gamma$. We define a value of this vector at the point where $dE_{tot}/d\gamma = 0$. The minimum of the function $E_{tot}(\gamma)$ corresponds to the ground state energy of the system in which the electrons are described by the trial wave function $\Phi_\gamma(\mathbf{r}_1, \mathbf{r}_2)$ being a product of the one-electron wave functions (7). The function $E_{tot}(\gamma)$ is given in Fig. 2. The total energy of the system, according to Fig. 2, reaches minimum at $\gamma_{min} \approx 0.4269$ and the ground state energy of the system is $E_{tot}(\gamma_{min}) \approx -1.589 \,\text{eV}$. The second electron affinity of $C_{60}$ is $\varepsilon_2 = -1.589 + 2.65 = 1.061 \,\text{eV}$; the first one is $\varepsilon_1 = 2.65 \,\text{eV}$.

Thus, assuming that the trial wave function of the pair of electrons has the form of a product of the one-electron wave functions (7), we found the electronic structure of the $C_{60}^{2-}$ anion by the variational method. It is known that the energy found by the variational equation for the ground state of the system cannot be less than an exact value. That is if the Dirac-bubble potential model is correct then the double charged negative ion exists in a stable state and the electronic level of the ground state in the real anion with two extra electrons is located deeper than $\varepsilon_2 = 1.061 \,\text{eV}$.

### 4. Variational method for the Lorentz-bubble potential model

The total energy in this case is divided into three parts: kinetic energy of electrons $E_{kin}$, potential energy of their interaction with the $C_{60}$ shell $E_{pot}$ described by potential (1) and electrostatic energy of electron interaction $Q$

$$E_{tot} = 2E_{kin} + 2E_{pot} + Q. \qquad (15)$$

Kinetic electron energy is [6]

$$E_{kin} = \frac{1}{2}\int_0^\infty \left(\frac{dR_{1s}}{dr}\right)^2 r^2 dr, \qquad (16)$$

where $R_{1s}(r)$ is the radial part of the trial wave function. The potential energy of electron interaction with $C_{60}$ shell is

$$E_{pot} = \int R_{1s}^2 U(r) r^2 dr. \qquad (17)$$

The parameters of the potential well $U(r)$ in (17) are presented in Table 1; we consider in this calculation the potential well thickness $\Delta=1$. For the trial wave functions of the pair of electrons, we choose a product of two one-electron wave functions $R_{1s}(r) = \chi_{10}(r)/r$ with two (Table 2) and three (Table 3) varied parameters. The functions $R_{1s}(r)$ are given in these Tables with accuracy of up to a normalization factor. In these formulas the following designations have been introduced: $r_>$ and $r_<$ represent $r>R$ and $r<R$, respectively. The behavior of the trial wave functions $R_{1s}(r)$ is qualitatively similar to the behavior of the dashed-line function in Fig. 1. The second electron affinity of $C_{60}$ from the last line in Table 2 is $\varepsilon_2$= -1.5794+2.65=1.0706 eV; extracted from the last line of Table 3 the value of energy is $\varepsilon_2$= -1.6518+2.65=0.9982 eV.

Table 2. Two-parameter trial wave functions

| Trial wave functions | $R$, au | $\beta$, au | $E_{tot}$, eV |
|---|---|---|---|
| $R_{1s}(r) \propto \exp[R\sqrt{r} - \beta r]$ | 9.7 | 2.0 | -1.0809 |
| $R_{1s}(r) \propto \exp[Rr - \beta r^2]$ | 1.36 | 0.11 | -1.1019 |
| $R_{1s}(r) \propto \exp[-(r-R)^2/\beta^2]$ | 6.07 | 3.30 | -1.1318 |
| $R_{1s}(r) \propto [\exp(-\beta R)/\beta R][\sinh(\beta r_<)/r_<]$, $R_{1s}(r) \propto [\sinh(\beta R)/\beta R][\exp(-\beta r_>)/r_>]$ | 6.658 | 0.272 | -1.4056 |
| $R_{1s}(r) \propto \cosh^{-1}[\beta(r-R)]$ | 6.27 | 0.54 | -1.4736 |
| $R_{1s}(r) \propto [(r-R)^2 + \beta^2]^{-1}$ | 6.30 | 2.30 | -1.5794 |

Table 3. Three-parameter trial wave functions

| Trial wave functions | $R$, au | $\beta_1$, au | $\beta_2$, au | $E_{tot}$, eV |
|---|---|---|---|---|
| $R_{1s}(r) \propto \exp[-(r_< - R)^2/\beta_1^2]$, $R_{1s}(r) \propto \exp[-(r_> - R)^2/\beta_2^2]$ | 6.13 | 3.7 | 3.2 | -1.1423 |
| $R_{1s}(r) \propto \{\exp[\beta_1(r-R)] + \exp[\beta_2(r-R)]\}^{-1}$ | 6.28 | 0.55 | 0.54 | -1.4745 |
| $R_{1s}(r) \propto \{\cosh[\beta_2(r-R)]\}^{-\beta_1}$ | 6.27 | 0.96 | 0.56 | -1.4809 |
| $R_{1s}(r) \propto [(r_< - R)^2 + \beta_1^2]^{-1}$, $R_{1s}(r) \propto [(r_> - R)^2 + \beta_2^2]^{-1}$ | 6.45 | 2.90 | 2.03 | -1.6518 |

**5. $C_{60}^{2-}$ ion photodetachment**

Additional important detailed information on the electronic structure of anions can be obtained by the methods of photoelectron spectroscopy. So, we calculate the photodetachment cross sections of the $C_{60}^{2-}$ anion near the photodetachment thresholds. Due to the Coulomb interaction of electrons the threshold behavior of the photodetachment cross section changes significantly. Because of the Coulomb repulsion among electrons, the cross section of the reaction $C_{60}^{2-} + \omega = C_{60}^- + e$ vanishes exponentially for $(\omega - J_1) \to 0$ (here $J_1 = \varepsilon_2$ is the photodetachment potential of the doubly-charged ion). Incidentally, the photodetachment cross section of the singly-charged negative ion near the threshold of the reaction $C_{60}^- + \omega = C_{60} + e$, according to the Wigner threshold law, is proportional to $(\omega - J_2)^{3/2}$ [5]. Here $J_2 = \varepsilon_1$ is the photodetachment potential of singly-charged ion.

We first consider the reaction wherein the doubly-charged ion $C_{60}^{2-}$ is transformed into a singly-charged negative $C_{60}^-$ ion. The radial part of the wave function of the optical electron in the initial state is defined by the function $\chi_{\beta 0}(r)$ of Eq.(7) wherein the wave vector is $\beta = \gamma_{\min}$; the potential energy of the $s$-level is equal to $J_1$. Here we calculate the photodetachment cross section in the "frozen core" approximation. Namely, the continuum wave functions are calculated in the undistorted field of the $C_{60}^-$ ion. The potential of this field is a sum of the bubble potential $U(r)$ and the potential (11) created by the charge of the electron residing in the potential well: $U(r) + W(r)$. The mean potential $W(r)$ is calculated using the wave function $\chi_{\gamma_{\min} 0}(r)$. In this approximation we neglect the changes in field of the $C_{60}^-$ ion caused by the photoelectron emission.

The radial parts of the continuum wave functions $\chi_{kl}(r)$ with the specific orbital angular momentum $l$ obey the wave equation

$$\frac{1}{2}\left[\frac{d^2\chi_{kl}}{dr^2} - \frac{l(l+1)}{r^2}\chi_{kl} + k^2\chi_{kl}\right] - [U(r) + W(r)]\chi_{kl} = 0. \tag{18}$$

Here $k^2/2 = \omega - J_1$ is the kinetic energy of the photoelectron. For large distances the potential $[U(r) + W(r)]_{r\to\infty} \to 1/r$ coincides with the Coulomb repulsion potential. Therefore, the wave functions $\chi_{kl}(r)$ at $kr \gg 1$ have the following asymptotic form

$$\chi_{kl}(r) \approx \sin\left[kr + \frac{Z}{k}\ln(2kr) - \frac{\pi l}{2} + \delta_l\right]. \tag{19}$$

Here the charge of the electron cloud formed by the extra electron of $C_{60}^-$ is $Z = -1$. The photoionization cross section for the process is determined by the formula

$$\sigma(\omega) = \frac{8\pi N_{10}}{3k}\alpha\omega |\langle k,1|d|10\rangle|^2. \tag{20}$$

In (20) $\alpha$ is the fine structure constant; $N_{10} = 2$ is the number of electrons in the potential well. The dipole matrix element in (20) is defined by the integral

$$\langle k,1|d|10\rangle = \int \chi_{k1} r \chi_{\gamma_{\min} 0} dr. \tag{21}$$

The calculated photodetachment cross section for the process $C_{60}^{2-} + \gamma = C_{60}^- + e$ is presented in Fig. 3. The cross section for the reaction $C_{60}^- + \omega = C_{60} + e$ is given in the same figure. In the latter case the potential $W(r)$ in equation (18) is equal to zero; the radial part of the electron wave function in the initial state is defined by the function $\chi_{\beta 0}(r)$ in Eq. (7). In the asymptotic region of (19) the charge $Z=0$ and the number of electrons is $N_{10} = 1$. Besides the photodetachment cross sections of the $C_{60}$ negative ions, the atomic negative ion photodetachment cross section calculated within the zero-range potential model [17] is given in the same figure. The dashed and dotted curves in Fig. 3 coincide with the cross sections calculated in paper [12]. According to this figure, the values of the disintegration cross sections of the fullerene anions $C_{60}^{z-}$ with $z$=1 and 2 are of the same order as the photodetachment cross sections of atomic ions with the same electron affinity.

### 6. Conclusions

It should be noted that the energy of the ground state of the $C_{60}^{2-}$ system calculated with the help of the variational equation cannot be less than the experimental value. The variational method as such cannot provide the location of the system ground level deeper than the experimental one. The essence of the approximation used in the paper was to select a trial wave function of the pair of electrons as a product of the one-electron wave functions. It was shown that already in this approximation the total energy of the system is negative. The next-order corrections can only increase the depth of level location corresponding to the ground state of the system. Consequently, within the Dirac- and Lorentz-bubble potential models the $C_{60}^{2-}$ anion is stable and the detachment energy of this anion (the second electron affinity energy) is equal to (or even less than) the value calculated in this paper ($\varepsilon_2$=0.9982 eV).

It was shown that near threshold behavior of the photodetachment cross sections $\sigma(\omega)$ is exhibit peculiar and interesting features. The first cross section accompanied by the transformation of the doubly-charged negative ion into a singly-charged one is exponentially small near the process threshold. The second cross section corresponds to the photodetachment of a singly-charged ion; it increases at the threshold as a power function of the kinetic energy of the photoelectron. According to given calculations these cross sections are of the same order as the photodetachment cross sections of atomic ions with the same electron affinity.

We hope that the data presented herein will prompt experimental works to look into the matter, thereby promoting such developments. Note that the long-lived ($\tau$>10$^{-3}$ sec) doubly charged carbon cluster anions $C_{60}^{2-}$ in the gas phase have been reported earlier from sputtering of a graphite surface by $Cs^+$ ions [18] and in experiments [7] on the laser desorption from a surface covered with neutral $C_{60}$ molecules.

### Acknowledgments

The author is very grateful to Dr. I. Bitenskiy for useful comments. This work was supported by the Uzbek Foundation Award Ф2-ФА-Ф164 (ASB) and U.S. DOE, Basic Energy Sciences, Office of Energy Research (AZM).

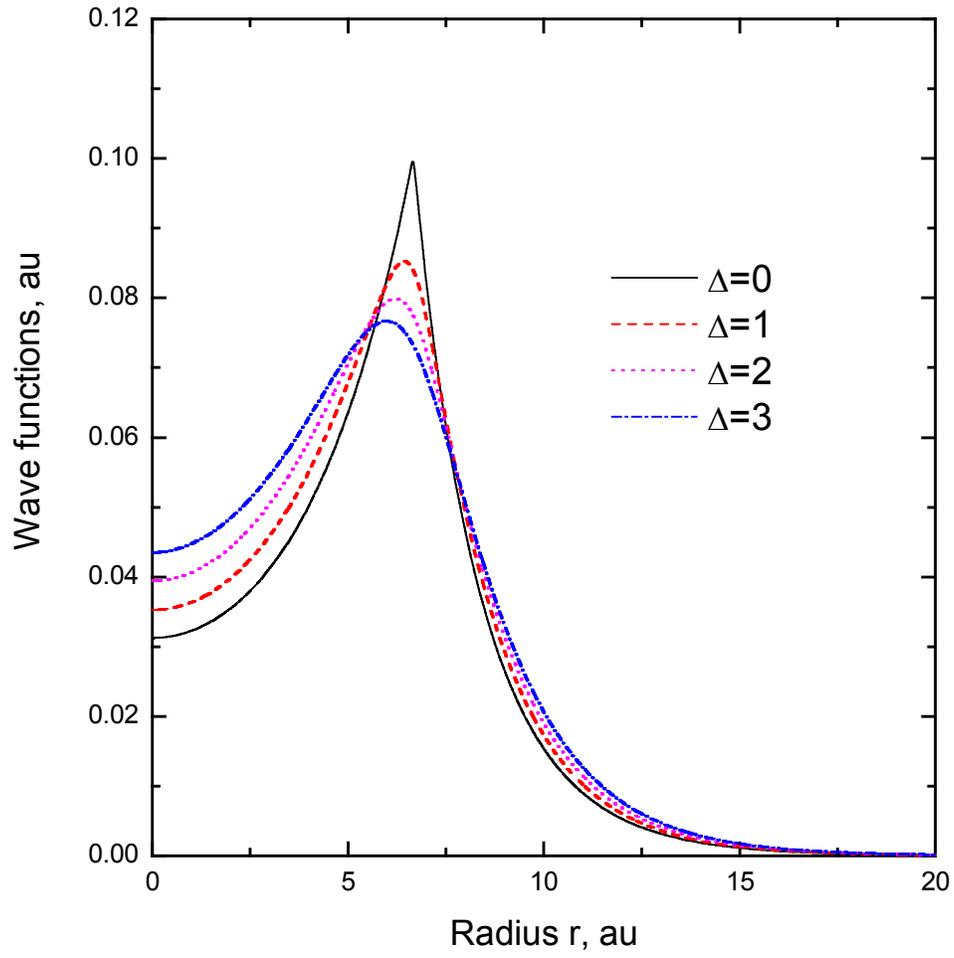

Fig. 1. Variation of the wave functions $\chi_{\beta 0}(r)/r$ with the different thicknesses, $\Delta$ of the Lorentz-bubble potential well

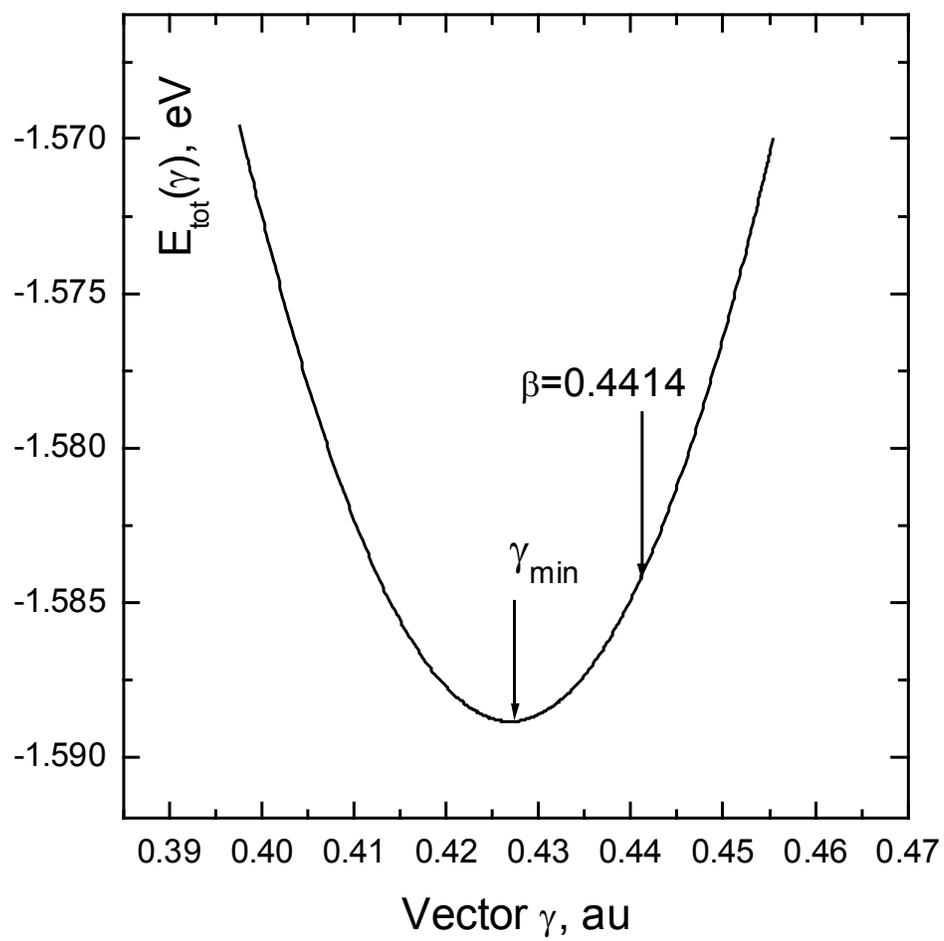

Fig. 2. The function $E(\gamma)$ as a function of $\gamma$

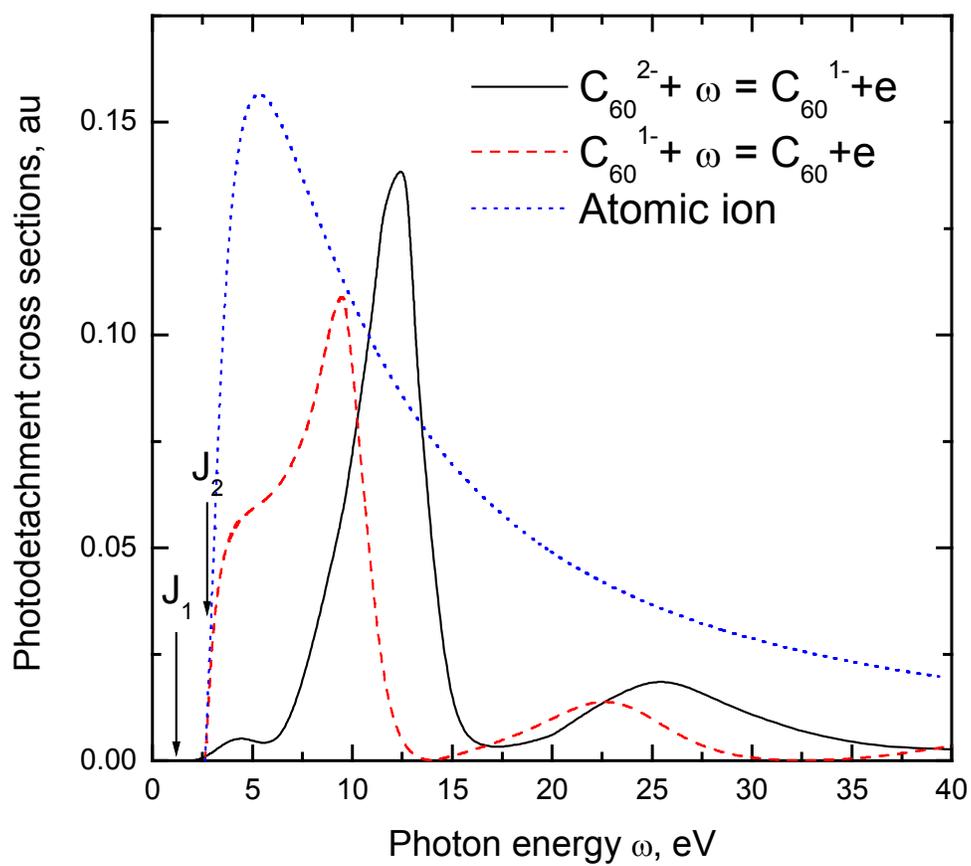

Fig. 3. Photodetachment cross sections versus photon energy, eV